\documentclass[12pt]{article}

\usepackage{amsmath, euscript, amssymb, amsfonts}

\newcommand{\R}{{\mathbb R}}
\newcommand{\V}{{\mathbb V}}
\newcommand{\C}{{\mathbb C}}
\newcommand{\T}{{\mathbb T}}
\newcommand{\w}{{\bf w}}
\newcommand{\st}{{\tt d}}
\newcommand{\di}{{\rm d}}
\newcommand{\sd}{{\mathfrak s}}
\newcommand{\W}[1]{{:\!#1\!\!:}}
\newcommand{\M}[1]{\sqrt{#1^2}}
\newcommand{\ew}{{\cal W}}

\newcommand{\dv}[2]{\langle #1,#2\rangle}

\begin{document}

\baselineskip=20pt

\begin{center}
{\Large ON WICK POWER SERIES CONVERGENT TO NONLOCAL FIELDS}

\bigskip
{\large
A.~G.~Smirnov\footnotemark{},
M.~A.~Soloviev\footnotemark[1]{}}\footnotetext{ Lebedev Physics Institute,
RAS, Moscow, Russia.  }
\end{center}

\vspace{0.3cm}

\begin{center}
{\large Abstract}
\end{center}

\bigskip
The infinite series in Wick powers of a generalized free field are
considered that are convergent under smearing with analytic test
functions and realize a nonlocal extension of the Borchers
equivalence classes. The nonlocal fields to which they converge
are proved to be asymptotically commuting, which serves as a
natural generalization of the relative locality of the Wick
polynomials. The proposed proof is based on exploiting the
analytic properties of the vacuum expectation values in $x$-space
and applying the Cauchy--Poincar\'e theorem.

\section{Introduction}
In this paper we continue the investigation~\cite{SS1,SS2}
of the infinite series of the form
\begin{equation}
\sum_{k=0}^{\infty} d_k\W{\phi^k}(x)
\label{1}
\end{equation}
in Wick powers of a neutral scalar field $\phi$, whose basic point
is the systematic use of the analytic properties of the vacuum
expectation values in $x$-space. The developed approach is
primarily aimed at applications to gauge field theory, where the
two-point function $w(x-x')=\langle\Psi_0,\phi(x)\phi(x')\Psi_0
\rangle$ does not necessarily satisfy the positivity condition
$w(f^*\otimes f)\geq 0$ (where $f$ is a test function), and it not
only allows easily finding the test function class on which a
given series is convergent, but also enables one to establish the
properties of the limiting field $\varphi$. In the positive metric
case, it is customary to use another approach~\cite{R} based on
estimating the terms of the series representing the vacuum
expectation value $\langle\Psi_0,\,
 \varphi(x_1)\ldots \varphi(x_n)\Psi_0\rangle$ in momentum space, where
they are expressible through positive measures. It is commonly
assumed that this is the only possible way of handling the problem
if the sum $\varphi$ of a series is a nonlocal field because in
this case the analyticity domain of its vacuum expectation values
in $x$-space is empty. Nevertheless, our approach is applicable to
such series as well if the analyticity properties of each their
particular term are duly taken into account, see~\cite{SS1}. It is
essential that this approach covers the fields $\phi$ of zero
mass, which were not considered in~\cite{R}, and, moreover, the
generalized free fields in a space-time of arbitrary dimension
$\st$. Here we shall prove the relative asymptotic commutativity
of the nonlocal fields to which the series in Wick powers of a
generalized free field converge. The role of the asymptotic
commutativity condition in the theory of nonlocal interactions was
analyzed in~\cite{S2}, where it was proved that this condition
ensures the normal connection between spin and statistics and the
CPT-invariance. Within the framework of the traditional approach,
the properties of nonlocal Wick series of a free field with
nonzero mass were considered earlier in~\cite{L}, whereas the
relation of the essential locality condition used in~\cite{L} to
the asymptotic commutativity is discussed in~\cite{S3}.  The
general construction of Wick powers of generalized free fields was
considered in~\cite{B}. Other important motivations for a deeper
analysis of nonlocal Wick series (in addition to the fact that
they form an extension of the Borchers equivalence class of the
field $\phi$) are the connection of nonlocal quantum field models
exhibiting singular ultraviolet behavior with string theory and
M-theory~\cite{K}, especially in the context of AdS/CFT
correspondence~\cite{G}, and the use of nonlocal formfactors for
removing ultraviolet divergences in phenomenological models
proposed as an alternative to string theory~\cite{M}. In
particular, the developed technique may be useful for the
treatment of the problem of a possible CPT-invariance breaking in
such models, which is discussed in~\cite{M}.

 \section{Analytic properties of two-point function}

From this point on, we shall assume $\phi$ to be a tempered
distribution generalized free field~\cite{BLOT}. The $n$-point
vacuum expectation values are expressible in terms of the
two-point one by the same recurrence relation as in the case of a
free field, which makes it possible to define the Wick ordered
powers $\W{\phi^k}(x)$. The analytic function whose boundary value
is the distribution $w(x)$ will be denoted by $\w(z)$. As is shown
in~\cite{SS1}, if the positivity condition is satisfied, the test
function class on which series~(\ref{1}) is convergent is
determined by the behavior of $\w(z)$ in the imaginary directions,
its growth for ${\rm Im}\,z\to 0$ (resp. for ${\rm Im}\,z\to
\infty$) being essential in the localizable (resp. nonlocalizable)
case. As a characteristic of this behavior one can take the
restriction of the function ${\bf w}(z)$ to the semi-axis ${\rm
Re}\,z=0$,
  ${\rm Im}\,z=(-\tau, 0,\dots, 0)$,  $\tau>0$, and we shall denote this restriction by
$u(\tau)$. The formula for the Laplace transformation $$\w(z) =
 (2\pi)^{-\mathrm d}\int\exp(-ipz) \hat w(p)\,\di p$$,
where $\hat w(p)$ is a positive polynomially bounded measure
supported by the closed upper light cone $\bar \V_+$, shows that
$u(\tau)$ is a strictly positive nondecreasing function which
majorizes $|{\bf w}(z)|$ for ${\rm Im}\,z$ of the specified form
and for any ${\rm Re}\,z$. Hence, by the Lorentz invariance of
$\hat w(p)$, we have
\begin{equation}
|\w(x+iy)|\leq u(\M{y}),\quad {\rm for\,\,all}\quad  y\in\V_-,
\label{2}
\end{equation}
where $y^2$ is the Lorentz square of $y$. Thus, $u(\tau)$ indeed
can serve as an indicator function according to the
definition~\cite{SS1}, where it was denoted by
$w_{{\scriptscriptstyle UV}}(\tau)$ in order to distinguish it
from the function characterizing the infrared behavior of $w$,
which is necessary in theories with an indefinite metric.
Moreover, the function $u(\tau)$ is the least one among all
functions satisfying (\ref{2}) and, therefore, is the best
characteristic of the behavior of $\w(z)$. It should also be noted
that $u(\tau)$ is infinitely differentiable and increases
indefinitely with decreasing argument.

As usual, we denote by $L_+(\C)$ the component of identity of the
complex Lorentz group and $\T^{ext}$ the extended analyticity
domain of $\w(z)$ which is generated from the primitive domain
$\R^{\st}+i\V_-$ by applying arbitrary transformations in
$L_+(\C)$ according to the Bargmann--Hall--Wightman theorem, whose
proof for an arbitrary space-time dimension can be found
in~\cite{J}.  The domain $\T^{ext}$ is invariant under the full
reflection $z\to -z$. Indeed, for an even $\st$ the reflection
belongs to $L_+(\C)$, and the general case can be treated as
follows. If $z$ is in $T^{ext}$, then there exists a
transformation $\Lambda\in L_+(\C)$ which takes $z$ to a point
with an imaginary part belonging to the negative $y^0$-semi-axis.
Now the statement follows if we note that the composition of
$\Lambda$ with the partial reflection
$(z^0,z^1,z^2\ldots,z^{\st-1})\to
(-z^0,-z^1,z^2\ldots,z^{\st-1})$, which also belongs to $L_+(\C)$,
takes $-z$ to a point with the same imaginary part. In particular,
the inclusion $\R^{\st}+i\V\subset \T^{ext}$ is valid. The
following simple lemma allows us to estimate the function $\w(z)$
for real arguments provided we know its behavior in the imaginary
directions.

{\bf Lemma 1.}
{\it
Let $0\leq\tau'<\tau$, $z=x+iy$, and let $x^2\leq -\tau^2$,
$y^2>-{\tau'}^2$. Then there exists
$\Lambda\in L_+(\C)$ such that $\Lambda z\in \R^{\st}+i\V_-$
and $\left(\mathrm{Im}\,
\Lambda z\right)^2\geq \tau^2-{\tau'}^2$.
}

{\bf Proof.} Suppose first that $y^2>0$ and let $\Lambda_1$ be a
real Lorentz transformation taking $y$ to a vector of the form
$(\tilde y^0,0,\ldots,0)$. Let $\Lambda_2$ be a pure rotation
which takes $\Lambda_1x$ to a vector $(\tilde x^0,\tilde
x^1,0,\ldots,0)$. Set $\Lambda=\Lambda_3\Lambda_2\Lambda_1$, where
$\Lambda_3(z^0,\ldots,z^{\st-1})=(\pm iz^1,\pm
iz^0,z^2,\ldots,z^{\st-1})$. Then $\mathrm{Im}\,\Lambda
z=(\pm\tilde x^1,\pm\tilde x^0,0,\ldots,0)$ belongs to $\V_-$
under the proper choice of the sign, and $(\mathrm{Im}\,\Lambda
z)^2=-x^2\geq\tau^2$. Now suppose $-{\tau'}^2<y^2\leq 0$. Then
there exists a Lorentz transformation $\Lambda_1$ taking $y$ to a
vector $y_1$ such that $\|{\bf y}_1\|\leq\tau'$. Let
$\Lambda_2,\Lambda_3$, and $\Lambda$ be defined as before, and let
$\tilde x=\Lambda_2\Lambda_1 x$ and $\tilde y=\Lambda_2\Lambda_1
y=\Lambda_2 y_1$. Then $\mathrm{Im}\,\Lambda z=(\pm\tilde
x^1,\pm\tilde x^0,\tilde y^2,\ldots,\tilde y^{\st-1})$,
$(\mathrm{Im}\,\Lambda z)^2\geq -x^2-\|\tilde {\bf
y}\|^2=-x^2-\|{\bf y}_1\|^2\geq \tau^2-{\tau'}^2$, and the proper
choice of the sign in the definition of $\Lambda_3$ ensures that
$\mathrm{Im}\,\Lambda z\in \V_-$.  The lemma is proved.

It is well known that all spacelike vectors belong to $\T^{ext}$,
which also follows from Lemma~1. For any such vector $x$ one can
find $\Lambda\in L_+(\C)$ such that $\Lambda x=-x$, and hence
$\w(x)=\w(-x)$. Therefore, by the uniqueness theorem,
\begin{equation}
\w(z)=\w(-z),\quad z\in \T^{ext}.
\label{3}
\end{equation}
In particular, at the level of the two-point vacuum expectation
values, the locality is a consequence of the other Wightman
axioms. Using the notation
\begin{equation}
G^{\tau}=\bigl\{x\in\R^{\st}\bigl|\,x^2\leq -\tau^2\bigr\}, \quad
V^{\tau}=\bigl\{y\in\R^{\st}\bigl|\,y^2>-\tau^2\bigr\}
\label{4}
\end{equation}
and combining~(\ref{2}) with Lemma~1, we obtain the estimate
\begin{equation}
|\w(z)|\leq u(\sqrt{\tau^2-{\tau'}^2}),\quad z\in G^{\tau}+i
V^{\tau'}\subset \T^{ext},
\label{5}
\end{equation}
which holds for $0\leq \tau'<\tau$.

\section{Wightman functions of Wick power series}

Let us denote the series~(\ref{1})by $\sd$. We shall also consider
its subordinate series and use the notation
$\sd'\triangleleft\,\sd$ which means that for all indices $k$,
with the possible exception of a finite subset the inequality
$|d'_k|\leq C_{\sd'}|d_k|$ holds, where $C_{\sd'}$ is a positive
constant. It is reasonable to impose the following conditions on
the coefficients of series~(\ref{1}):
\begin{equation}
 d_k\ge 0, \quad
d_k\,d_l\le Ch^{k+l}d_{k+l},
 \label{6}
\end{equation}
where $C$,$h$ are constants whose role is explained in~\cite{SS1};
the subordinate series need not satisfy them. We use the
Gelfand--Shilov spaces $S^b$ as test function spaces,
see~\cite{GS}. The defining index $b$ can be regarded as an
indicator function characterizing the momentum space behavior of
the test functions. More precisely, the derivatives of the Fourier
transform of $f\in S^b(\R^n)$ satisfy the inequalities
\begin{equation}
 |\partial^{\,\kappa} \hat
f(p)|\,b\left(\frac{|p|}{B}\right)\leq C_\kappa,
 \label{7}
 \end{equation}
with $C_\kappa$ and $B$ positive constants depending on $f$. For
definiteness, we assume the norm $|\cdot|$ in $\R^n$ to be
uniform. In the context of QFT, the function $b$ characterizes the
high energy behavior of the fields defined over $S^b$.

Let $\varphi_{\sd_1},\ldots, \varphi_{\sd_n}$ be the fields
determined by the series $\sd_j\triangleleft\,\sd$.  Consider the
$n$-point vacuum expectation value ${\cal
W}_{\sd_1,\ldots,\sd_n}(x_1,\ldots,x_n)=
\dv{\Psi_0}{\varphi_{\sd_1}(x_1)\ldots
\varphi_{\sd_n}(x_n)\Psi_0}$. Applying the Wick theorem gives the
well-known formal representation
\begin{equation}
 \ew_{\sd_1,\ldots,\sd_n}= \sum_K D_K W^K,
 \label{9}
\end{equation}
where $K$ is an integer-valued multi-index with nonnegative
components $k_{j,m}$, $1\leq j< m\leq n$, which have the sense of
the number of pairings between the terms of the series $\sd_j$ and
$\sd_m$, and the designation
\begin{equation}
 W^K=
\prod_{1\leq j<m\leq n}w(x_j-x_m)^{k_{jm}}
\label{10}
\end{equation}
is used. The numerical coefficients $D_K$ are expressible in terms
of the coefficients of the series $\sd_j$ in the following way:
 \begin{equation}
D_K=\frac{\kappa!}{K!}\prod_{1\leq j\leq
n}d^{(j)}_{\kappa_j},\quad {\rm where} \quad
\kappa_j=k_{1j}+\ldots+k_{j-1,j}+k_{j,j+1}+\ldots+k_{jn}.
\label{11}
\end{equation}
If the distribution series on the right-hand side of~(\ref{9})
absolutely converges on each test function in the space
$S^b(\R^{n\tt d})$, then it is unconditionally summable with
respect to the strong topology of its dual space because the
latter is a Montel space. If this is the case for any set of
series subordinate to $\sd$, then, as is shown in~\cite{SS1}, the
fields $\varphi_{\sd'}$, ${\sd'\triangleleft\,\sd}$ are well
defined as operator-valued generalized functions over $S^b(\R^{\tt
d})$ acting in the Hilbert space ${\cal H}$ of the initial field
$\phi$. In particular, the vector series that define the repeated
action of these operators on the vacuum $\Psi_0$ are
unconditionally convergent, and the linear span of all vectors of
the form $\varphi_{\sd_1}(f_1)\ldots \varphi_{\sd_n}(f_n)\Psi_0$,
$f_i\in S^b(\R^{\tt d})$ serves as a common dense invariant domain
of definition for the family of fields
$\{\varphi_{\sd'}\}_{\sd'\triangleleft\,\sd}$ in ${\cal H}$. We
shall denote this domain by $D(\sd)$.

The distribution~(\ref{10}) is the boundary value of the analytic function
 \begin{equation}
 \mathbf W^K(z)=\prod_{1\leq j<m\leq n}\w(z_j-z_m)^{k_{jm}}
\label{12}
\end{equation}
from the cone $$V_{n-} = \{y\in\R^{n\tt d}\,|\, y_j-y_m\in\V_-,\,
1\leq j<m\leq n\},$$ and the representation~(\ref{9}) can be
rewritten in the following more precise form:
\begin{equation}
 \ew_{\sd_1,\ldots,\sd_n}(x_1,\ldots,x_n)= \sum_K D_K {\bf b}_{V_{n-}} {\bf
 W}^K,
 \label{13}
\end{equation}
where ${\bf b}_{V_{n-}}$ is the boundary value operator. It is
worth noting that the function~(\ref{12}) is defined and analytic
in the open set $T_n=\{z\in\C^{n\st}\,|\,z_j-z_m\in
\T^{ext},\,j\ne m\}$ and, in particular, in the tube $\R^{n\st}+i
V_n$, where $V_n=\{y\in\R^{n\st}\,|\,y_j-y_m\in \V,\,j\ne m\}$.
For what follows, it is essential to know the transformation law
for representation~(\ref{13}) under the rearrangements   of the
operators $\varphi_{\sd_j}(x_j)$ entering into the vacuum
expectation value.

{\bf Lemma 2.}
{\it
Suppose the distribution series on the right-hand side of $(\ref{13})$
is unconditionally convergent in $S^b(\R^{n\tt d})$ for any
$\sd_j\triangleleft\,\sd$, $j\leq n$.
Let $\pi$ be a permutation of the indices $(1,\ldots,n)$.
Then
\begin{equation}
\ew_{\sd_{\pi 1},\ldots,\sd_{\pi n}}(x_{\pi
1},\ldots,x_{\pi n})= \sum_K D_K {\bf b}_{\pi V_{n-}} {\bf W}^K,
\label{14}
\end{equation}
where $\pi V_{n-} =
\{y\in\R^{n\tt d}\,|\, y_{\pi j}-y_{\pi m}\in\V_-,\, 1\leq j<m\leq n\}$.}

{\bf Proof.}
Because of (\ref{13}) we have
\begin{equation}
\ew_{\sd_{\pi 1},\ldots,\sd_{\pi n}}(
x_{\pi 1},\ldots,x_{\pi n})=
\sum_K D'_K  ({\bf b}_{V_{n-}} {\bf W}^K)(x_{\pi1},\ldots,x_{\pi n}),
\label{15}
\end{equation}
where $D'_K$ is the coefficient corresponding, by~(\ref{11}), to
the permuted set $\sd_{\pi 1},\ldots,\sd_{\pi n}$. Let $K'$ be the
multi-index whose components $k'_{jm}$ are equal to $k_{\pi j,\pi
m}$ for $\pi j<\pi m$ and $k_{\pi m,\pi j}$ for $\pi j>\pi m$.
From~(\ref{12}), it follows that ${\bf
W}^{K'}(z_{\pi1},\ldots,z_{\pi n})$ coincides with the product
defining ${\bf W}^K(z)$ to within the signs of the arguments of
some factors, and in view of (\ref{3}) we conclude that ${\bf
W}^K(z_1,\ldots,z_n)={\bf W}^{K'}(z_{\pi1},\ldots,z_{\pi n})$ in
$T_n$. Passing in this equality to the boundary values from the
cone $\pi V_{n-}$, we obtain $({\bf b}_{\pi V_{n-}} {\bf
W}^K)(x_1,\ldots,x_n)=({\bf b}_{V_{n-}} {\bf
W}^{K'})(x_{\pi1},\ldots,x_{\pi n})$, whereas the relation
$\kappa'_{i}=\kappa_{\pi i}$, which follows from (\ref{11}),
implies that $D'_{K'}=D_K$. Making the change $K\to K'$ of the
summation indices in the unconditionally convergent
series~(\ref{15}) and applying the above identities, we arrive
at~(\ref{14}). The lemma is proved.

 \section{Generalization of the locality axiom}

If the space $S^b$ on which series~(\ref{1}) converges contains
functions of compact support, i.e., the field $\varphi_{\sd}$ is
an (operator-valued) ultradistribution, then the fulfilment of the
Wightman axioms for this field is easily established by the same
arguments as in~\cite{SS2}, where even the more general case of an
indefinite metric was considered. In particular, the locality of
$\varphi_{\sd}$ and, moreover, the relative locality of the fields
$\varphi_{\sd'}$, ${\sd'\triangleleft\,\sd}$, immediately follow
from the relative locality of the Wick monomials $\W{\phi^k}(x)$.
This property can be also derived from Lemma~2 if $\pi$ is taken
to be the transposition $\tau_j$ of the neighbouring indices $j$
and $j+1$. The distribution ${\bf b}_{V_{n-}} {\bf W}^K -{\bf
b}_{\tau_jV_{n-}} {\bf W}^K$ is supported by the closed cone
\begin{equation}
{\bar V}_{j,j+1}=\{x\in \R^{n\tt d}\,|\,(x_j-x_{j+1})^2\geq 0\}.
\label{16}
\end{equation}
Therefore, the support of the functional
\begin{equation}
\ew_{\sd_1,\ldots,\sd_n}(x_1,\ldots,x_j,x_{j+1},\ldots,x_n)-
\ew_{\,\sd_1,\ldots,\sd_{j+1},\sd_j,\ldots,\sd_n}
(x_1,\ldots,x_{j+1},x_j,\ldots,x_n)
\label{17}
 \end{equation}
is also contained in this cone, whence
$[\varphi_{\sd_j}(f_j),\varphi_{\sd_{j+1}}(f_{j+1})]\Psi=0$ for
spacelike separated supports of the test functions $f_j,
f_{j+1}\in S^b$ and for any $\Psi\in D(\sd)$ because $n$ and the
rest of $\sd_i$ in~(\ref{17}) can be taken arbitrary.

The locality condition implies that the vacuum expectation values
in momentum space have less than exponential growth, see,
e.g.,~\cite{BLOT}. For this reason, the Gelfand--Shilov space
determined by the indicator function $b(s)=e^s$, which is
customarily denoted by $S^1$, is universal for local fields. The
elements of $S^1$ allow analytic continuation into a complex
neighbourhood of the real space and never have compact support.
Nevertheless, the methods of the hyperfunction theory make it
possible to give a correct definition of support for the
functionals belonging to the dual space $S^{\prime 1}$. We refer
the reader to~\cite{S5} for details and confine ourselves to
saying that the support of $v\in S^{\prime 1}$ is contained in a
closed cone $K$ if and only if $v$ has a continuous extension to
each space $S^1(U)=\bigcup_{B>0}S^{1,B}(U)$, where $U$ is an open
cone containing $K\setminus\{0\}$ and $S^{1,B}(U)$ consists of
analytic functions on the complex $1/B$-neighbourhood $\tilde
U^{1/B}$ of the set $U$ with the property that the norms
$$\|f\|_{U,N,B} = \sup_{z\in\tilde U^{1/B}}|f(z)|(1+|x|)^N$$ are
finite. The corresponding generalization of local commutativity
for the fields $\varphi$, $\varphi'$ defined over $S^1(\R^{\tt
d})$ means that the matrix elements
$\dv{\Phi}{[\varphi(x),\varphi'(x')]\Psi}$ have support in the
closed cone
\begin{equation}
\bar V^{(2\st)}=\{(x, x')\in \R^{2\st}\,|\,(x-x')^2\geq 0\}
\label{18a}
\end{equation}
for any $\Phi$ and $\Psi$ belonging to the common domain of
definition of these fields. Within the framework of hyperfunction
theory, Wick power series of the free field of nonzero mass were
studied earlier in the work~\cite{NM}. In our notation, the
restriction on the series coefficients found there takes the form
\begin{equation}
\lim_{k\to\infty}(d_{2k}\,k!)^{1/k}= 0.
\label{19}
\end{equation}
The fulfilment of the Wightman axioms for the (hyper)fields
defined by Wick series was established in~\cite{NM} indirectly, by
means of an equivalence theorem of Osterwalder--Schrader type for
a properly modified Euclidean field theory and the Minkowski
quantum field theory formulated in terms of
Fourier-hyperfunctions. Below we shall obtain, as a by product, a
simple direct proof showing that in this case the limiting fields
satisfy the generalized locality condition.

If $\ln b(s)$ grows faster than linearly, then the elements of
$S^b$ are entire functions and the functionals belonging to
$S^{\prime b}$ are nonlocal. However, under suitable restrictions
on $b$, they inherit an important part of the properties of
hyperfunctions which has the sense of angular localizability.
In~\cite{S5} a corresponding theory has been developed for the
spaces whose indicator functions are exponentials of order $>1$,
and in~\cite{SS2} it has been extended to a more general case. In
the work~\cite{SS2}, another scale of spaces $S^b_a$ was
considered which is required for the indefinite metric field
theory, but the construction proposed there is applicable to $S^b$
as well. Namely, let $\beta(\tau)$ be a nonnegative, convex,
differentiable, and indefinitely increasing function on the
half-axis $\tau\geq 0$ and let $U$ be an open cone in $\R^n$. We
denote by $\delta_U(x)$ the distance from the point $x$ to the
cone $U$ and consider the space ${\cal
E}^{\beta}(U)=\bigcup_{B>0}{\cal E}^{\beta,B}(U)$, where ${\cal
E}^{\beta,B}(U)$ consists of entire analytic functions on $\C^n$
such that the norms
\begin{equation} \|f\|_{U,N,B} =
\sup_{x,y}|f(x+iy)|(1+|x|)^N
\exp\{-\beta(B|y|)-\beta\circ\delta_U(Bx)\}
\label{20}
\end{equation}
are finite for any $N=0,1\ldots$. The topology of ${\cal
E}^{\beta}(U)$ is defined to be that of the inductive limit of the
countably normed spaces ${\cal E}^{\beta,B}(U)$ with the index
$B\to \infty$.

{\bf Lemma 3.}
{\it The space ${\cal E}^{\beta}(\R^n)$ coincides with the space
$S^b$ defined by the indicator function
$b(s)=e^{\beta_*(s)}$, where $\beta_*(s)=\sup_{\tau>0}(s\tau-\beta(s))$.}

{\bf Proof.}
Taking into account the elementary inequalities
$1+|z|\leq (1+|x|)(1+|y|)$ and
  $\beta((1+\epsilon)\tau)- \beta(\tau)\geq
C_\epsilon+h_\epsilon\tau$, where $\epsilon>0$ is arbitrarily
small and $h_\epsilon$ is a positive constant, we see that
replacing the factor $(1+|x|)^N$ in~(\ref{20}) by
$\max_{|\kappa|\leq N}|z^\kappa|$ leads to an equivalent
definition of ${\cal E}^{\beta}$. Let $f\in {\cal
E}^{\beta}(\R^n)$. Then $$
 |\partial^{\,\kappa} \hat f(p)|=\left|\int_{{\rm
Im}z=y}\!z^ke^{ipz}f(z)\,\di z\right|\leq
'_\kappa\int\!(1+|x|)^{-(n+1)}e^{-py+\beta(B|y|)}\,\di x. $$
Making use of the freedom in the choice of the plane of
integration, we set $y=\tau p/|p|$ and take the infimum with
respect to $\tau>0$. As a result, we obtain an estimate of the
form~(\ref{7}) with $b=e^{\beta_*}$. Conversely, if (\ref{7})
holds for such an indicator function, then taking the inverse
Fourier--Laplace transformation, we find that $$ |z^\kappa
f(z)|\leq (2\pi)^{-n}\int\!e^{n|p| |y|}|\partial^{\,\kappa} \hat
f(p)| \,\di p \leq C_{\kappa,\epsilon}\sup_p e^{n|p|
|y|-\beta_*((1-\epsilon)|p|/B)}, $$
and so $f\in {\cal
E}^{\beta}(\R^n)$ because the Legendre transformation is
involutory. The lemma is proved.

The spaces $S^b$ of the specified type will be called the
Gelfand--Shilov--Gurevich spaces because they also belong to the
class of spaces of type $W$ introduced by B.~L.~Gurevich. Among
such spaces, a special role is played by that defined by
$\beta(\tau)=\tau$. This space customarily denoted by $S^0$ is
nothing but the Fourier transformed Schwartz's space $\cal D$. It
is universal for nonlocal fields because in this case $b(s)=1$ for
$0\leq s\leq 1$ and $b(s)=\infty$ for $s>1$, i.e., the test
functions have compact support in momentum space, and so fields
with an arbitrarily singular ultraviolet behavior can be smeared
with them.

{\bf Definition 1. }
A closed cone $K$ is called a carrier cone of a functional
$v\in {{\cal E}'}^{\beta}$ if $v$ has a continuous extension to each space
${\cal E}^\beta(U)$, $U\supset K\setminus\{0\}$.

For any $v\in {{\cal E}'}^{\beta}$ there exists a unique minimal
closed carrier cone. This has been proved in~\cite{S5} for the
spaces $S^b$ defined by the exponentials of order $>1$, and just
the properties of them that were used in this proof are included
into the definition of ${\cal E}^{\beta}$.

{\bf Definition 2. } We say that the fields $\varphi$,  $\varphi'$
defined on the test function space ${\cal E}^{\beta}(\R^{\tt d})$
asymptotically commute for large spacelike separations of their
arguments if the matrix elements
$\dv{\Phi}{[\varphi(x),\varphi'(x')]\Psi}$ are carried by
cone~$(\ref{18a})$ for any $\Phi,\Psi$ in the common domain of
definition of these fields.

 \section{Conditions of convergence of Wick power series on analytic
test functions}

A general convergence criterion for Wick series can be formulated
in terms of the above-mentioned characteristic $u(\tau)$ of the
two-point function $\w(z)$ and the indicator function $b(s)$ as
follows. The series~(\ref{1}) is convergent under smearing with
test functions in $S^b$ if
\begin{equation}
\sum_{k=0}^{\infty} L^kk!\,d_{2k}\, \inf_{\tau>0}\,u(\tau)^k e^{s\tau}
\le C_{L,\epsilon}\, b(\epsilon s)
\label{8}
\end{equation}
for any $L>0$ and $\epsilon>0$. The proof of this criterion is the
same as that of Theorem~4 in~~\cite{SS1}, where it has been
established in the case of the free field of mass $m$ and the
explicit form of the corresponding function $u(\tau)$ has been
used. For the Gelfand--Shilov--Gurevich spaces, an alternative
formulation is possible in terms of the function $\beta$, which
will be useful below. Its derivation is much simpler than that in
the general case considered in~\cite{SS1}, where test functions
are not necessarily analytic.

 {\bf Theorem 1.} {\it Let $\phi$ be a scalar neutral generalized free
 field and let
$$ u(\tau)=(2\pi)^{-\mathrm d}\int\! e^{-\tau p^0}\hat w(p)\,\di
p, $$ where $\hat w$ is the Fourier transform of its two-point
function. Suppose restrictions~$(\ref{6})$ on the coefficients of
the Wick power series $\sd$ hold. If the function $\beta$ defining
the space ${\cal E}^{\beta}$ satisfies the condition
\begin{equation}
\sum_{k=0}^{\infty} L^k k!\,d_{2k}
 \,\inf_{\tau>0}u(\tau)^{k}e^{\beta(B\tau)} < \infty
 \label{21}
\end{equation}
for arbitrarily large $L,B>0$, then the field $\varphi_{\sd}$ and
all fields $\varphi_{\sd'}$, ${\sd'\triangleleft\,\sd}$, are well
defined as operator-valued generalized functions over ${\cal
E}^{\beta}(\R^{\tt d})$ acting in the Hilbert space ${\cal H}$ of
the initial field $\phi$.}

{\bf Proof.} According to what has been said above, it is
sufficient to show that (\ref{21}) implies the absolute
convergence of the series on the right-hand side of (\ref{9}) on
every test function $f\in{\cal E}^{\beta}$. Let
$\eta=(\tau,0,\ldots,0)\in \V_+$ and let
$y=(\eta,2\eta,\ldots,n\eta)\in V_{n-}$. Then
\begin{equation}
W^K(f) =
\int\!\mathbf W^K(x+iy)f(x+iy)\,\di x.
\label{lab12}
\end{equation}
Making use of (\ref{2}), (\ref{20}), the monotonicity of $u$, and
the equality $|\eta|=\M{\eta}=\tau$, we obtain the estimate
 $$
 |\mathbf W^K(x+iy)f(x+iy)| \leq
\frac{\|f\|_{2n\st,B}}{(1+|x|)^{2n\st}} u(\tau)^{|K|}
e^{\beta(Bn\tau)},\quad f\in {\cal E}^{\beta,B}(\R^{n\st}).
 $$
In view of the freedom in the choice of $\tau$ we have
$$
|W^K(f)|\leq C(f)\inf_{\tau>0}u(\tau)^{|K|}e^{\beta(Bn\tau)}.
$$
Thus, the required convergence of the series~(\ref{9}) is ensured by the
convergence of the number series
\begin{equation}
\sum_K |D_K|\,\inf_{\tau>0}u(\tau)^{|K|}e^{\beta(B\tau)}
\label{lab13}
\end{equation}
with arbitrarily large $B>0$. From the conditions~(\ref{6}) and
the properties of the polynomial coefficients, it follows that
$|D_K|\leq h^{\prime {|K|}}\,|K|!\,d_{2|K|}$, see~\cite{SS1}.
Since the number of multi-indices with the norm $|K|=k$ does not
exceed $k^{n(n-1)/2}$, we conclude that series~(\ref{lab13}) is
majorized by series~(\ref{21}) for sufficiently large $L$. The
theorem is thus proved.

{\bf  Lemma 4.} {\it Criterion~$(\ref{8})$ is equivalent to
condition~$(\ref{19})$ for $b(s)=e^s$ and condition~$(\ref{21})$
for $b(s)=e^{\beta_*(s)}$.}

{\bf Proof.} Let $b(s)=e^s$. Under condition~(\ref{19}) the series
$\sum d_{2k}\,k! z^k$ is convergent everywhere. Majorizing the
infimum on the left-hand side of inequality~(\ref{8}) by the value
of the function at $\tau=\epsilon$, we see that it is valid with
the constant $$C_{L,\epsilon}=\sum d_{2k}\,k!(Lu(\epsilon))^k.$$
To prove the inverse implication (\ref{8}) $\Rightarrow$
(\ref{19}), choose $m>0$ such that $$C=(2\pi)^{-\mathrm
d}\int_{\bar\V_+\cap\{p\,:\,p^0\leq m\}}\hat w(p)\,\di p>0.$$ Then
$u(\tau)\geq Ce^{-m\tau}$ and condition~(\ref{8}) ensures that
$$(C L)^k\,|d_{2k}|\,k!\leq C_{L,\epsilon}e^{\epsilon s}$$ for
$mk\leq s$. Setting $s=mk$ and making use of the arbitrariness of
$L$, we obtain~(\ref{19}).

Now let $b(s)=e^{\beta_*(s)}$. If $\beta$ grows linearly, then the
statement of the lemma is verified immediately. So we assume that
$$\lim_{\tau\to\infty}\beta(\tau)/\tau=\infty$$. Let us
demonstrate that
\begin{equation}
\inf_{\tau>0}u(\tau)^{k}e^{\beta(\tau)}=
\sup_{s\geq 0}\inf_{\tau>0} u(\tau)^{k}e^{s\tau-\beta_*(s)}.
\label{20a}
\end{equation}
(This relation generalizes the conclusion of Lemma~1
in~\cite{SS2}, where $u(\tau)=1/\tau$.) Since the Legendre
transformation is involutory, the equality~(\ref{20a}) holds for
$k=0$. Let $k>0$. Since $u(\tau)\geq Ce^{-m\tau}$, the infimum on
the left-hand side occurs at some finite point $\tau_k>0$
satisfying the equation $\beta'(\tau_k)=-k u'(\tau_k)/u(\tau_k)$.
Set $s_k=\beta'(\tau_k)$. Making use of the relation $$
\frac{\di^2\ln u(\tau)}{\di\tau^2}= \frac{1}{2
u(\tau)^2}\int_{\R^{\mathrm d}\times \R^{\mathrm d}}
(p^0-p^{\prime 0})^2 e^{-(p^0+p^{\prime 0})\tau}\hat w(p)\hat
w(p')\,\di p\di p'\geq 0 $$ and taking into account that, for
convex functions, every stationary point is the point of absolute
minimum, we conclude that the extrema $\inf_{\tau>0}
e^{-s_k\tau+\beta(\tau)}=e^{-\beta_*(s_k)}$ and $\inf_{\tau>0}
u(\tau)^{k}e^{s_k\tau}$ are attained at $\tau=\tau_k$. Estimating
from below the supremum with respect to $s$ by the value of the
function at the point $s_k$, we obtain $$ \sup_{s\geq
0}\inf_{\tau>0}u(\tau)^{k} e^{s\tau-\beta_*(s)}\geq
e^{-s_k\tau_k+\beta(\tau_k)}\inf_{\tau>0}u(\tau)^{k} e^{s_k \tau}
= e^{\beta(\tau_k)}u(\tau_k)^{k}. $$ Since $\inf_{\tau}\sup_s
G(\tau,s)\geq \sup_s\inf_{\tau}G(\tau,s)$ for any function $G$,
the inverse inequality also holds, and so (\ref{20a}) is proved.
Supposing (\ref{8}) is valid and applying (\ref{20a}), we have
$$L^k d_{2k} k!\,\inf_{\tau>0}u(\tau)^{k}e^{\beta(B\tau)}\leq
C_{2L,1/B}\,2^{-k},$$ whence (\ref{21}) immediately follows.
Conversely, setting $B=1/\epsilon$ in (\ref{21}), using
(\ref{20a}) and estimating from below the suprema with respect to
$s$ by the value of the function at a fixed point, we arrive at
(\ref{8}). The lemma is proved.

{\raggedright
   \section{Proof of asymptotic commutativity}}

We proceed to show that the generalizations of local commutativity
considered in Section~4 are fulfilled for the fields determined by
the Wick series convergent on analytic test functions.

{\bf Theorem 2.} {\it Let  $\sd$ be a series in the Wick powers of
a generalized free field $\phi$ and let its coefficients satisfy
assumption~$(\ref{6})$. Suppose the indicator function $b(s)$ of
the test function space satisfies condition~$(\ref{8})$. If $S^b$
is a Gelfand--Shilov--Gurevich space, then the fields
$\varphi_{\sd'}$ determined by the series
${\sd'\triangleleft\,\sd}$ commute asymptotically. If $S^b=S^1$,
then they are relatively local in the sense of hyperfunction
theory.}

{\bf Lemma 5}. {\it Let the condition~$(\ref{21})$ be satisfied.
If for any set of series
 $\sd_j\triangleleft\,\sd$, $1\leq j\leq n$,
functional~$(\ref{17})$ has a continuous extension to the space
 ${\cal E}^{\beta}(V_{j,j+1})$, where $V_{j,j+1}$ is the open cone
$\{x\in \R^{n\tt d}\,|\,(x_j-x_{j+1})^2>0\}$, then all the fields
determined by series subordinate to $\sd$ commute asymptotically.
If condition~ $(\ref{19})$ holds and the functional~$(\ref{17})$
allows a continuous extension to $S^1(V_{j,j+1})$, then they are
relatively local in the sense of hyperfunction theory.}

{\bf Proof.} Let
$\Phi=\varphi_{\sd_1}(f_1)\ldots\varphi_{\sd_l}(f_l)\Psi_0$ and
$\Psi=\varphi_{\sd'_1}(g_1)\ldots\varphi_{\sd'_m}(g_m)\Psi_0$,
where $\sd_i,\sd'_k\triangleleft\sd$, $f_i,g_k\in {\cal
E}^{\beta}(\R^{\tt d})$. Then the value of the functional
$\langle\Phi,[\varphi_{\sd'}(x'),\varphi_{\sd''}(x'')]\Psi\rangle$
on a test function $f\in {\cal E}^{\beta}(\R^{2\st})$ coincides
with that of the functional of the form~$(\ref{17})$ with a
suitable set of indices on the test function $\bar
f_l\otimes\ldots\otimes\bar f_1\otimes f\otimes
g_1\otimes\ldots\otimes g_m$. If $f\in {\cal
E}^{\beta}(V^{(2\st)})$, then this tensor product is the element
of ${\cal E}^{\beta}(V_{j,j+1})$ which depends continuously on
$f$. Thus, the existence of a continuous extension to ${\cal
E}^{\beta}(V_{j,j+1})$ for the functionals of the form~(\ref{17})
ensures that the matrix elements in question have continuous
extensions to ${\cal E}^{\beta}(V^{(2\st)})$ and all the more to
${\cal E}^{\beta}(U)$, where $U\supset \bar
V^{(2\st)}\setminus\{0\}$. Analogous statements, with proper
changes in notation, are valid for the space $S^1$. Since $D(\sd)$
is the linear span of the vectors $\Phi,\Psi$ of the specified
form, the lemma is proved.

In the above derivation of the convergence criterion~(\ref{21})
the key role is played by the variation of the plane of
integration in the representation~(\ref{lab12}) which enables us
to obtain the best estimate for each term of the series on the
right-hand side of (\ref{9}). We shall apply the same idea to
prove the asymptotic commutativity of the sums of Wick series.
However, this will require integrating the corresponding analytic
functions over surfaces of a more complicated form. The variation
of such surfaces in the analyticity domain is admissible by the
Cauchy--Poincar\'e theorem~\cite{V}, but for obtaining concrete
estimates it will be convenient to use directly the Stokes theorem
which lies at the basis of its derivation.

According to Lemma~2 functional~(\ref{17}) is represented by the
series
\begin{equation}
\sum_K D_K(\mathbf b_{V_{n-}} {\bf W}^K
-\mathbf b_{\tau_jV_{n-}} {\bf W}^K).
\label{lab17}
\end{equation}
In view of the barrelledness of ${\cal E}^{\beta}(V_{j,j+1})$ and
$S^1(V_{j,j+1})$, in order to extend continuously this functional
to these spaces it is sufficient to construct a continuous
extension of each term of series~(\ref{lab17}) and show that it is
absolutely convergent on every element of the corresponding space.
We shall consider in detail such a procedure for ${\cal
E}^{\beta}(V_{j,j+1})$ and then explain how the proof should be
modified for the case of hyperfunctions.

For $\tau\geq 0$ and $1\leq j\leq n$, we define the regions
\begin{eqnarray*}
&&\Gamma_{j}^{\tau}=\bigl\{x\in
\R^{n\st}\,\bigl|\,(x_j-x_{j+1})^2\leq-\tau^2\bigr\} = \bigl\{x\in
\R^{n\st}\,\bigl|\,x_j-x_{j+1}\in G^{\tau}\bigr\},\\
&&U_{j}^{\tau} =
\bigl\{y\in
\R^{n\st}\,\bigl|\,y_i-y_l\in \V\quad \mbox{{\rm for
}}i<l,\,(i,l)\ne(j,j+1); \quad y_j-y_{j+1}\in V^\tau\bigr\}.
\end{eqnarray*}
The relations (\ref{2})--(\ref{5}) show that, for
$0\leq\tau'<\tau$, the set $\Gamma_{j}^{\tau}+iU_{j}^{{\tau}'}$
lies in the analyticity domain of $\mathbf W^K(z)$, and for
$z=x+iy\in \Gamma_{j}^{\tau}+iU_{j}^{{\tau}'}$, the following
inequality holds:
\begin{equation}
|\mathbf W^K(z)|\leq
u\left(\sqrt{\tau^2-{\tau'}^2}\right)^{K_{j,j+1}}
\prod_{i<l,\,(i,l)\ne(j,j+1)}u\left(\M{(y_i-y_l)}\right)^{K_{il}}.
\label{lab18}
\end{equation}

{\bf Lemma 6.} {\it Let $\tau>0$, $\eta=(\tau,0,\ldots,0)\in
\V_+$, and $f\in {\cal E}^{\beta}(\R^{n\st})$. With the notation
\begin{eqnarray}
&&y(t)
=(\eta,\ldots,(j+(1-t)/2)\eta,(j+(1+t)/2)\eta,(j+2)\eta,\ldots,n\eta),
\nonumber\\ &&F(x,t) = \mathbf W^K(x+iy(t))f(x+iy(t)),
\label{lab26}
\end{eqnarray}
for any $K$ and $1\leq j\leq n$, the identity
\begin{equation}
(\mathbf b_{V_{n-}} {\bf W}^K -\mathbf b_{\tau_jV_{n-}} {\bf
W}^K)(f) = I_1+I_2 \label{lab23}
\end{equation}
holds, where
\begin{eqnarray}
I_1 &=& \int_{\complement\Gamma_{j}^{\tau}}(F(x,1)-F(x,-1))\,\di
x, \label{lab24}\\ I_2 &=&i\tau\int_{-1}^1 \di
t\int_{\R^{(n-1)\mathrm d}} \di \xi_1\ldots \di \xi_{j-1}\di
\xi_{j+1}\ldots \di \xi_n \int_{\partial G^{\tau}} F(P^{-1} \xi,t)
\nu^0\,\di S_{\tau}, \label{lab25}
\end{eqnarray}
$P$ is the linear operator taking $x=(x_1,\ldots,x_n)$ to
$(x_1-x_2,\ldots,x_{n-1}-x_n,x_n)$, $\nu$ is the unit inward
normal to $\partial G^{\tau}$, and $\di S_{\tau}$ is the surface
measure on $\partial G^{\tau}$.}

{\bf Proof.} Note that $y(1)\in V_{n-}$ and $y(-1)\in \tau_j
V_{n-}$ and hence $$(\mathbf b_{V_{n-}} {\bf W}^K -\mathbf
b_{\tau_jV_{n-}} {\bf W}^K)(f) = \int_{\R^{n\mathrm
d}}(F(x,1)-F(x,-1))\,\di x.$$ We split the integration domain into
$\Gamma_{j}^{\tau}$ and $\complement\Gamma_{j}^{\tau}$, rewrite
the right-hand side of the last equality as $I_1+J$, where $$J =
\int_{\Gamma_{j}^{\tau}}(F(x,1)-F(x,-1))\,\di x,$$ and shall show
that $J=I_2$. Let $0<\tau'<\tau$. From (\ref{20}), (\ref{lab18}),
it follows that, for any compactum $Q\subset U_{j}^{{\tau}'}$, the
estimate $$|\mathbf W^K(z)f(z)|\leq C_{N,Q} (1+|x|)^{-N}$$ holds
if $z\in \Gamma_{j}^{\tau}+i Q$. Making use of Cauchy's integral
formula, one can easily show that analogous inequalities are
satisfied for the derivatives as well. Therefore, the function
$$\Phi(y)= \int_{\Gamma_{j}^{\tau}}\mathbf W^K(x+i y)f(x+iy)\,\di
x$$ is differentiable and, applying the Cauchy--Riemann equations,
we have $$\frac{\partial \Phi(y)}{\partial y^0_l} =
i\int_{\Gamma_{j}^{\tau}} \frac{\partial [\mathbf
W^K(x+iy)f(x+iy)]}{\partial x^0_l}\,\di x,\quad y\in
U_{j}^{{\tau}'}.$$ Since $y(t)\in U_{j}^{{\tau}'}$ for $-1\leq
t\leq 1$ and $$\int_{\Gamma_{j}^{\tau}}F(x,t)\di x=\Phi(y(t)),$$
we obtain $$ J=\int_{-1}^1 \di t \frac{\di}{\di t}\Phi(y(t)) =
\frac{i\tau}{2}\int_{-1}^1 \di t
\int_{\Gamma_{j}^{\tau}}\left(\frac{\partial F} {\partial
x_{j+1}^0}(x,t) - \frac{\partial F}{\partial x_j^0}(x,t)\right)
\,\di x. $$ Let us denote $\tilde F(\xi,t)= F(P^{-1} \xi,t)$ and
make the change of variables $x \to P^{-1}\xi$. Then, observing
that $P(\Gamma_{j}^{\tau})=\R^{(n-1)\st}\times G^{\tau}$, we get
the equality $$ J = \frac{i\tau}{2}\int_{-1}^1 \di t
\int_{\R^{(n-1)\mathrm d} \times G^{\tau}} \left( \frac{\partial
\tilde F}{\partial \xi_{j+1}^0}(\xi,t) + \frac{\partial \tilde
F}{\partial \xi_{j-1}^0}(\xi,t) -2\frac{\partial \tilde
F}{\partial \xi_j^0}(\xi,t)\right) \,\di \xi. $$ From (\ref{20}),
(\ref{lab18}), it follows that $\tilde F(\xi,t)\to 0$ as $|\xi|\to
\infty$, $\xi\in \R^{\st (n-1)}\times G^{\tau}$, and therefore the
integration of $\partial \tilde F/\partial \xi_{j+1}$ and
$\partial \tilde F/\partial \xi_{j-1}$ yields zero. Reducing the
multiple integral to the iterated one, we obtain $$J =
-i\tau\int_{-1}^1 \di t \int_{\R^{(n-1)\mathrm d}}\di\hat
\xi\int_{G^{\tau}}\di \xi_j \frac{\partial \tilde F}{\partial
\xi_j^0}(\xi,t),$$ where $\di\hat \xi=\di \xi_1\ldots \di
\xi_{j-1}\di \xi_{j+1}\ldots \di \xi_n$. The last integrand can be
regarded as the divergence of the vector field $(\tilde
F,\mathbf{0})$. Applying the Stokes theorem and replacing $\tilde
F(\xi,t)$ by $F(P^{-1}\xi,t)$, we conclude that $J=I_2$.

{\bf Lemma 7.} {\it The functional $\mathbf b_{V_{n-}} {\bf W}^K
-\mathbf b_{\tau_jV_{n-}} {\bf W}^K$ can be continuously extended
to ${\cal E}^{\beta}(V_{j,j+1})$ and, for any $B>0$, $N\geq
(n+1)\st$ there exist $B',C$ such that
\begin{equation}
|(\mathbf b_{V_{n-}} {\bf W}^K -\mathbf b_{\tau_jV_{n-}} {\bf
W}^K)(f)|\leq C
\|f\|_{V_{j,j+1},N,B}\inf_{\tau>0}u(\tau)^{|K|}e^{\beta(B'\tau)}
\label{lab26a}
\end{equation}
for all $f\in {\cal E}^{\beta,B}(V_{j,j+1})$ }

{\bf Proof.} Let us show that formulas (\ref{lab23})-(\ref{lab25})
determine the desired extension if $f$ is assumed to be an element
of the space ${\cal E}^\beta(V_{j,j+1})$. We first estimate $I_1$
supposing $f\in {\cal E}^{\beta,B}(V_{j,j+1})$. It is easy to see
that $\delta_{V_{j,j+1}}(x)\leq \tau$ for $x\in
\complement\Gamma_{j}^{\tau}$. Besides, $|y(t)|\leq n|\eta|=n\tau$
for $-1\leq t \leq 1$ and, in view of (\ref{20}) and the
monotonicity of $\beta(s)$, we have
\begin{equation}
|f(x+iy(t))|\leq \|f\|_{V_{j,j+1},N,B}(1+|x|)^{-N}e^{2\beta(B
n\tau)},\quad x\in \complement\Gamma^{\tau}_{n,j}. \label{lab30}
\end{equation}
Taking into account that $y(\pm 1)\in V_n$ and using relations
(\ref{2}), (\ref{3}), the monotonicity of $u$, and the equality
$\M{\eta}=\tau$, we find that $$|F(x,\pm 1)|\leq
\|f\|_{V_{j,j+1},N,B}\,
(1+|x|)^{-N}u(\tau)^{|K|}e^{2\beta(Bn\tau)}$$ for
$x\in\complement\Gamma_{j}^{\tau}$. Thus, we have
\begin{equation}
|I_1|\leq C_1\,\|f\|_{V_{j,j+1},N,B}\,u(\tau)^{|K|}
e^{2\beta(Bn\tau)}. \label{lab31}
\end{equation}
Now let us estimate $I_2$.  Observe that if $\xi_j\in\partial
G^{\tau}$, then $P^{-1}\xi\in\partial \Gamma_{j}^{\tau}$.
Furthermore, $(1+|P|)( 1+|P^{-1}\xi|)\geq (1+|\xi|)$, where
$|P|=\sup_{|x|\leq 1}|P x|$. Using~(\ref{lab30}) and the
definition $|\xi|= \max (|\xi_j|,|\hat \xi|)$, we obtain
\begin{equation}
|f(P^{-1}\xi+iy(t))|\leq \frac{\|f\|_{V_{j,j+1},N,B}(1+|P|)^N}
{(1+|\hat \xi|)^{N-\st}(1+|\xi_j|)^{\st}}e^{2\beta(B n\tau)},\quad
\xi_j\in \partial G^{\tau}. \label{lab32}
\end{equation}
Next we apply (\ref{lab18}), taking into account that $y(t)\in
U_{j}^0$ for $t\ne 0$ and using the monotonicity of $u$. As a
result, we get
\begin{equation}
|\mathbf W^K(P^{-1}\xi+iy(t))|\leq u(\tau)^{|K|},\quad
\xi_j\in\partial G^{\tau}. \label{lab33}
\end{equation}
Further, there exists a constant $C$ independent on $\tau$ and
such that $$\int (1+|x|)^{-\st}|\nu^0|\,\di S_{\tau} \leq C.$$
Indeed,
\begin{equation}
|\nu^0|\,\di S_{\tau}=r^{\st-2} \,\di r \,\di\Omega, \label{lab28}
\end{equation}
where $r=\|\mathbf{x}\|$ and $\di\Omega$ is the area element for
the surface of the unit sphere in $\R^{\st-1}$.

Let us denote $$C_2= C (1+|P|)^N\int \frac{\di\hat \xi}{(1+|\hat
\xi|)^{N-\st}}$$ and substitute relations~(\ref{lab32}),
(\ref{lab33}) into (\ref{lab26}), (\ref{lab25}). As a result, we
obtain
\begin{equation}
|I_2|\leq C_2\,\tau\,\|f\|_{V_{j,j+1},N,B} u(\tau)^{|K|}
e^{2\beta(B n\tau)}. \label{lab34}
\end{equation}
The presence of the factor $\|f\|_{V_{j,j+1},N,B}$ in
estimates~(\ref{lab31}), (\ref{lab34}) ensures that formulas
(\ref{lab23})-(\ref{lab25}) define a continuous extension of the
distribution $\mathbf b_{V_{n-}} {\bf W}^K -\mathbf
b_{\tau_jV_{n-}} {\bf W}^K$ to ${\cal E}^{\beta}(V_{j,j+1})$ and,
by the Cauchy--Poincar\'e theorem, the extensions corresponding to
different $\tau$ coincide with each other. Since $\beta$ is
unbounded from above and convex, we have $C_1+C_2\tau\leq
C'\exp(\beta(Bn\tau))$, where $C'>0$. Besides, $3\beta(n
B\tau)\leq \beta(3n B\tau)+2\beta(0)$ because of the convexity of
$\beta$. Combining these inequalities with (\ref{lab31}),
(\ref{lab34}) and passing to the infimum with respect to $\tau$,
we obtain (\ref{lab26a}) with $B'=3n B$. The lemma is proved.

From Lemma~7, it immediately follows that series~(\ref{lab17}) is
absolutely convergent on every test function in ${\cal
E}^{\beta}(V_{j,j+1})$ provided number series~(\ref{lab13})
converges for any $B>0$. As was established in proving Theorem~1,
this is ensured by condition~(\ref{21}) which is equivalent to the
criterion~(\ref{8}) by Lemma~4. Thus, Theorem~2 is proved for the
case of spaces ${\cal E}^{\beta}$.

The proof of Lemma~6 is extended immediately to the case $f\in
S^{1,B}(\R^{n\st})$, if we assume $\tau<1/(nB)$. Repeating, with
appropriate changes, the derivation of Lemma~7, we make sure that,
for all $\tau<1/(nB)$, formulas (\ref{lab23})-(\ref{lab25}) define
the same continuous extension of the functional $\mathbf
b_{V_{n-}} {\bf W}^K -\mathbf b_{\tau_jV_{n-}} {\bf W}^K$ to
$S^{1,B}(V_{j,j+1})$ and
\begin{equation}
|(\mathbf b_{V_{n-}} {\bf W}^K -\mathbf b_{\tau_jV_{n-}} {\bf W}^K)(f)|\leq C
\|f\|_{V_{j,j+1},N,B} u\left(\frac{1}{nB}\right)^{|K|}.
\label{lab35}
\end{equation}
The extensions corresponding to different $B$ are obviously
compatible and so define a continuous extension to
$S^1(V_{j,j+1})$. Because of (\ref{lab35}), series (\ref{lab17})
is absolutely convergent on every $f\in S^1(V_{j,j+1})$ if $\sum_K
L^{|K|}\,|D_K|<\infty$ for all $L>0$. The estimate of the
coefficients $D_K$ mentioned in the proof of Theorem~1 shows that
the latter is ensured by the condition~(\ref{19}), and it remains
to apply Lemma~4 to complete the proof.

\section{Concluding remarks}

Together with the results of work~\cite{SS2} Theorem~2 shows that
the nonlocal fields determined by series in Wick powers of a
generalized free field satisfy all requirements of Wightman's
formulation if the locality axiom is replaced by the asymptotic
commutativity condition. It is noteworthy that using the
generalized Gelfand--Shilov spaces enables one to consider also
the series convergent on quasianalytic test function classes
defined by the indicator functions $b$ that grow slower than any
linear exponential but do not satisfy the strict localizability
condition $$\int_1^\infty \frac{\beta(s)}{s^2}{\rm d}s<\infty$$
ensuring that $S^b$ contains functions of compact support. In
fact, Theorem~2 is applicable to this case as well because the
generalized functions defined on such spaces have the same
supports as their restrictions to $S^1$, see~\cite{H}.

\noindent {\bf Acknowledgments.} This work was supported in part
by the Russian Foundation for Basic Research under Grants No.
99-01-00376, 99-02-17916, and 00-15-96566, and in part by INTAS
Grant No. 99-1-590.

\end{document}